\documentclass{atelier_egc}

\usepackage[latin1]{inputenc}
\usepackage{amsmath,graphicx}

\newcommand{\ds}{\displaystyle}
\newcommand{\be}{\begin{equation}}
\newcommand{\ee}{\end{equation}}
\newcommand{\bd}{\begin{displaymath}}
\newcommand{\ed}{\end{displaymath}}

\newcommand{\ba}{\begin{eqnarray}}
\newcommand{\ea}{\end{eqnarray}}
\newcommand{\ban}{\begin{eqnarray*}}
\newcommand{\ean}{\end{eqnarray*}}

\newcommand{\R} {I\!\!R}
\newcommand{\E} {I\!\! E}

\def\xt{\tau}
\def\bt{\boldsymbol{\xt}}
\def\bx{\boldsymbol{X}}

\renewcommand{\Box}{\hfill\rule{0.25cm}{0.25cm}} 

\newtheorem{Def}{Definition}[section]
\newtheorem{Prop}{Proposition}[section]

\newenvironment{dem}{\ \\ {\bf Proof. }}
{\Box\par\medskip\noindent}

\def\1{{\bf 1}}

\resume{Nous proposons une nouvelle méthode pour estimer les ruptures du
  rythme cardiaque dans les bandes parasympathiques et orthosympathiques basée
  sur la transformée en ondelettes dans le domaine complexe et l'étude des
  ruptures des moments des modules de ces transformées en ondelettes. Nous
  observons des ruptures dans la distribution pour les deux bandes de
  fr\'equence.}

\summary{We propose a new method for estimating the change-points of heart
  rate in the orthosympathetic and parasympathetic bands, based on the
  wavelet transform in the complex domain and the study of the change-points
  in the moments of the modulus of these wavelet transforms. We observe
  change-points in the distribution for both bands. }
\titrecourt{Detection of Change--Points in the Spectral Density}
\nomcourt{Bertrand, Teyssi\`ere, Boudet, Chamoux}
\titre{ Detection of Change--Points in the Spectral Density.\\ With
  Applications to ECG Data}

\auteur{Pierre R. Bertrand, Gilles Teyssière, Gil Boudet, Alain Chamoux }
\affiliation{INRIA Saclay, APIS Team
 \\UMR CNRS 6620, Clermont-Ferrand University,  France.
\\E-mail: Pierre.Bertrand@inria.fr,\\
\affilsep
CREATES, Aarhus University, Denmark 
\\E-mail: stats@gillesteyssiere.net\\
\affilsep
Institut de Médecine du travail, UFR Médecine,  Clermont-Ferrand University\\
Gil.Boudet@wanadoo.fr\\
\affilsep
Institut de Médecine du travail, UFR Médecine,  Clermont-Ferrand University\\ CHU Clermont-Ferrand
\\E-mail: alain.chamoux@u-clermont1.fr}

\begin{document}

\section{Introduction}

ECG signal analysis has a long story after the implementation of the
ambulatory monitoring by Holter at the beginning of the fifties.
Recent measurement methods, due to the size reduction of the measurement
devices, see Chamoux (1984), allow us to record ECG  data for
healthy people over a long period of time: long distance (marathon) runners,
individuals daily (24 hours) records,  etc. We then obtain large datasets that
allow us to characterize the variations of the heartbeat rate in the two
components of the nervous autonomous system: the parasympathetic and the
orthosympathetic ones.

\begin{figure}[ht!]
\begin{center}
\includegraphics[height=6cm,width=16cm]{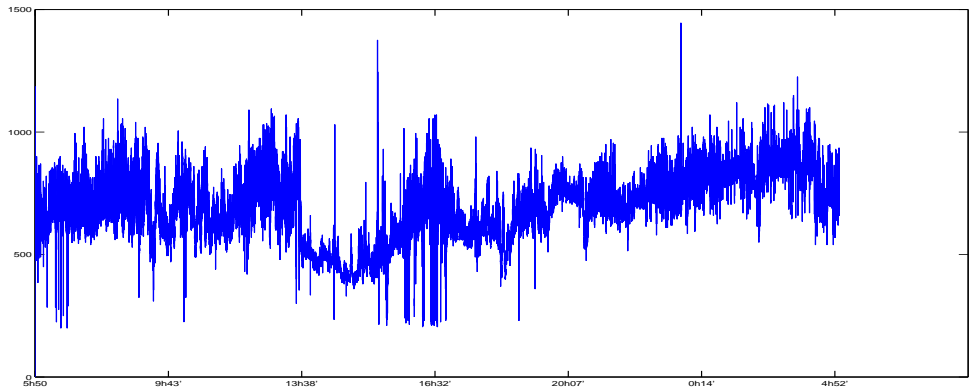}
\end{center}
\caption{RR interval for a healthy subject during a period of 24 hours}
\end{figure}
The data studied in this paper have been recorded the 29$^{th}$ of January
2008, at the Clermont-Ferrand University Hospital (CHU) by G. Boudet. Unlike
Boudet {\em  et al.} (2004)  and Cottin {\em et al} (2006), these data are not
recorded in the framework of a  laboratory experiment, but during the real
life of the individual under investigation.

We have daily datasets,  over 100,000 observations, and in the near future we
will have data over several days, i.e., over that sample size. We are
measuring the interval between two RR peaks, i.e., if we denote by
$(t_i)_{i=1,\dots,N}$,
the sequence of peak times measured with a precision of 1.E-03 second, we
consider the series $X(t_i) = \Big(t_i- t_{i-1}\Big)$ measured in seconds.
Several indicators are related to heartbeat data. The most popular is the
instantaneous average frequency, i.e., $X(t_i)^{-1}$, which is displayed by
runners watches. This quantity is informative on daily activity of observed
individual: sleeping and waking up times, physical activity, e.g., sport, manual
work, but does not summarize all relevant information. Note that heartbeat
data display large variations, clustering, etc, as only individuals with
serious disease display a regular heart rate.

Cardiologists are interested in the study of this signal in two frequency
bands: the orthosympathetic and parasympathetic bands, i.e., the frequency bands
$(0.04\,Hz,\,0.15\, Hz)$ and $(0.15\, Hz,\,0.5\, Hz)$ respectively.
The definition of these bands is the outcome of research works, see e.g., Task
force of the European Society of Cardiology and the North American Society of
Pacing and Electrophysiology (1996), and is based on
the fact that the energy contained inside these bands would be a relevant
indicator on the level of the stress of an individual.

Indeed, for the heart rate, the parasympathetic system is often compared to
the brake while the orthosympathetic system would be a nice accelerator; see
e.g. Goldberger (2001). At rest
there is a permanent braking effect on the heart rate. Any solicitation of the
cardiovascular system, any activity initially produces a reduction of
parasympathetic brake followed by a gradual involvement of the sympathetic
system. These mechanisms are very interesting to watch in many diseases
including heart failure, but also rhythm disorders that may fall under one or
other of these two effects, monitoring the therapeutic effect of several
Medicines including some psychotropic. In the field of physiology such data
are crucial for measuring the level of vigilance and particularly the risk of
falling asleep while driving a vehicle, the level of stress induced by
physical activity or level of perceived stress, which can be considered as a
criterion of overtraining in sport.

Fractals models have been used in cardiology after the works by Ivanov {\em et
  al.} (1999), who applied the multifractal spectrum analysis advocated by
Frisch (1996), for modeling RR series and classifying individuals according
to this multifractal spectrum, as this spectrum discriminates between
individuals who experienced hearth trouble, and those who did not. However,
this tool has some shortcomings as it requires huge samples common in
turbulence analysis, and is then inappropriate for studying phenomena
occurring at a resolution lower than the daily time interval, such as the
variations of the parasympathetic and orthosympathetic systems inside the day.

Wavelets based methods have been used in biostatistics by Diab {\em et al.}
(2007) for uterine EMG signal analysis. However, they consider that the
process studied is homogeneous, and used these methods as a classification
tool. Another significant difference is the fact that they use discrete
wavelet decomposition, i.e., a frequency decomposition on a dyadic wavelet
basis, the choice for the frequency bands is made without reference to a
biological phenomenon. In our case, the choice for the frequency band is
justified by biological considerations, and we fit the wavelets inside these
bands.

This is why the continuous analysis of both systems and their quantification
is a particularly promising research area. So, the example in Figures 4, 5, 6
shows at  the observation 28,220 i.e.,  13h40'50'', a simultaneous variation
of both systems.

\section{Mathematical formulation}
For biological reasons, the heart rate is within the interval $[20 ,
250]$ bpm (beats per minute), i.e.,  $X(t)$ belongs to the RR interval
$60/250 s.<X(t)<60/20 s.$. This leads to modeling by stationary or locally stationary processes.
We wish to decompose this signal in a sequence of homogeneous intervals
having the same mean and/or variance. We make the assumption that the series
is Gaussian, and then find the optimal segmentation of the process using the
same approach as in Lavielle and Teyssi\`ere (2006) or Lavielle and Moulines (2000).

We assume that the signal is the sum of a piecewise constant function and a
Gaussian process, centered and locally stationary. We then have the following
representation:

\begin{equation}
\label{repr:harmonizable} X(t)= \mu(t)+\int_{\R}
e^{it\xi}  f^{1/2}(t,\xi) \, dW(\xi), \quad \mbox{for all}\quad t  \in R,
\end{equation}
where
\begin{itemize}
  \item  $\xi\mapsto f(t,\xi)$ is an even and positive function, called spectral
    density piecewise constant, i.e., there exists a partition

$\tau_1,\dots, \tau_K$ such that  $f(t,\xi) = f_k(\xi)$ for $t\in [\tau_i, \tau_{i+1}[$
  \item the function $t\mapsto \mu(t)$ is also piecewise constant for another partition
  $\widetilde{\tau}_1,\dots, \widetilde{\tau}_{L}$ with $\mu(t) = \mu_{\ell}$ if $t\in[\widetilde{\tau}_{\ell},\,\widetilde{\tau}_{\ell+1} [$
\end{itemize}
The wavelet coefficient associated with $\psi$ is
\begin{equation}
W_\psi(b) = \int_{\R} \psi(t-b) X(t)\, dt\hspace{1cm} \mathrm{unit\;in}\; second^2.
\end{equation}
 In Bardet and Bertrand (2007) or  Bardet {\it et al.} (2008), one
 can find a  theoretical study of the wavelet coefficient for  stationary (or
 with stationary increment) centered Gaussian processes, i.e., for $X$ given
 by (\ref{repr:harmonizable}) with $\mu(t) = 0$. This approach can be
 generalized to locally stationary Gaussian processes, this will be detailed
 in a forthcoming paper.

According to recommendations of Task force of the European Soc. Cardiology and the North American Soc.
of  Pacing and Electrophysiology (1996), 
we use the following notations:
\begin{itemize}
\item $[\omega_1,\omega_2] = (0.04\,Hz,\,0.15\, Hz)$ denotes the
  orthosympathetic frequency band,
\item  $[\omega_2,\omega_3] =(0.15\, Hz,\,0.5\, Hz)$ denotes the
  parasympathetic band
\end{itemize}
The energy associated with each of these frequency bands and localized around
the time $b$, is measured by the modulus of the complex wavelet coefficients
$|W_i(b)|^2$ for $i=1, 2$, with
 \ban
b\longmapsto W_i(b)&=&  \int_{\R} \psi_i(t-b)\,X(t)\,dt, \quad i=1,2.
\ean
In this work, these wavelets coefficients are computed at each second, i.e.,
the difference between two consecutive values for $b$ is equal to 1 second.
\subsubsection*{How to choose the wavelets  $\psi_1$ and $\psi_2$ ?}
In the idealistic case, one would use two filters $\psi_1$ and $\psi_2$ with
compact support, the Fourier transforms of which have support
inside the orthosympathetic and parasympathetic bands.

Unfortunately, it does not exist a non null function $\psi$ with compact
time domain support and compact frequency support, see for instance Mallat (1998)
Th 2.6 p.34. Therefore, the best we can do is to choose between a filter with
a compact frequency support and a filter with a compact time domain support. The
first choice is well suited for stationary models, see  Bardet and Bertrand
(2007).
But, in this work, we are interested by locally stationary models, thus our
specifications are a filter with compact time domain support as a Daubechies wavelet.
The price  to pay for the compactness of the time domain support is the loss in
the compactness of the frequency support.
However, the frequency support is {\em "almost compact"} 
in the following sense:
\begin{Def}[$\rho$ pseudo support] Let $0<\rho<1$, be a  map $g\in L^2(\R)$
  that admits the compact interval $I$ as a $\rho$ pseudo support if
$\,\ds\frac{\int_{I} |g(t)|^2\,dt }{\int_{\R} |g(t)|^2\,dt}=\rho.$
\end{Def}
Fourier transform of Daubechies wavelet have a reasonably small $\rho$ pseudo
support with a ratio $\rho$ close to $1$. Moreover, the larger the number of
the Daubechies wavelet is, the closer to $1$ the  ratio $\rho$  is; see
the example below with the Daubechies wavelet D6.

By scaling and modulation, one can adjust the pseudo support inside a
specified frequency band as stated by the following proposition
\begin{Prop}\label{Prop:scaling:modulation}
Let $\psi$ be a filter with compact support $[L_1,L_2]$ and a frequency  $\rho$ pseudo support
$[\Lambda_1,\Lambda_2]$, for any frequencies band $[\omega_1,\omega_2]$, the map
$\,\ds\psi_1(t) = \mu\times e^{i\eta t}\,\psi(\lambda t)$ with
\begin{eqnarray*}
\mu>0,\qquad\lambda=\frac{\omega_2-\omega_1}{\Lambda_2-\Lambda_1}& \mathrm{and}&  \eta=\frac{\omega_1+\omega_2}{2}-(\omega_2-\omega_1)\frac{\Lambda_2+\Lambda_1}{\Lambda_2-\Lambda_1}
\end{eqnarray*}
has a $\rho$ pseudo support
$[\omega_1,\omega_2]$ and a time domain support
$\ds \left[\frac{\Lambda_2-\Lambda_1}{\omega_2-\omega_1}L_1,\frac{\Lambda_2-\Lambda_1}{\omega_2-\omega_1}L_2\right]$.
\end{Prop}
\begin{dem} Since $\ds\widehat{\psi_1}(\xi) = \mu\times\widehat{\psi}\left(\frac{\xi-\eta}{\lambda}\right)$, one can deduce
$\;\ds\rho\; pseudo\, supp\, \psi_1= \eta+\lambda\times\rho\; pseudo\; supp\, \psi\;$ and then the proposition.
\end{dem}
\subsubsection*{Different choices for the wavelets  $\psi_1$ and $\psi_2$}
From Proposition \ref{Prop:scaling:modulation}, one can deduce the different possible choices of the filters  $\psi_1$ and $\psi_2$
\begin{itemize}
  \item  {\bf Daubechies wavelet D6:} In this case, we have $\Lambda_1=0.08$,
    $\Lambda_2=1.75$, $\rho=0.9999$ and one can set
 \ban \psi_1(t) = \mu\times e^{i\eta_1 t}\,D_6(\lambda_1 t)\quad \mathrm{and}
 \quad\ds\psi_2(t) = \mu\times e^{i\eta_2 t}\,D_6(\lambda_2 t)\\
 \mathrm{with} \quad\eta_1= -0.0255,\lambda_1= 0.0659, \eta_2= -0.0585,\;\mathrm{and}\;\lambda_2=  0.2096.
 \ean
  The length of the time support are $\left|Supp\, \psi_1\right|$ and
  $\left|Supp\, \psi_2\right| $.
  One can see on Fig. 2 (a) that the Fourier transforms $\widehat{\psi}_1(x)$ and $\widehat{\psi}_2(x)$ have almost disjoint supports.

  \ban\lambda_1=\frac{\omega_2-\omega_1}{\Lambda_2-\Lambda_1}& \mathrm{and}&  \eta_1=\frac{\omega_1+\omega_2}{2}-(\omega_2-\omega_1)\frac{\Lambda_2+\Lambda_1}{\Lambda_2-\Lambda_1}\\
  \lambda_2=\frac{\omega_3-\omega_2}{\Lambda_2-\Lambda_1}& \mathrm{and}&  \eta_2=\frac{\omega_2+\omega_3}{2}-(\omega_3-\omega_2)\frac{\Lambda_2+\Lambda_1}{\Lambda_2-\Lambda_1}\\
   \left|Supp\, \psi_1\right|= \frac{\Lambda_2-\Lambda_1}{\omega_2-\omega_1}\times\left|Supp\, D_6\right| & \mathrm{and}& \left|Supp\, \psi_2\right|= \frac{\Lambda_2-\Lambda_1}{\omega_3-\omega_2}\times\left|Supp\, D_6\right|\ean
  \item {\bf Gabor wavelet:} For computational reasons, we will use the Gabor wavelet defined as
\begin{equation}\label{def:Gabor}
\psi(t) = e^{i\eta t} g(t), \quad g(t) = \frac{1}{(\sigma^2 \pi)^{1/4}}\, e^{- \frac{t^2}{2\sigma^2}}
\end{equation}
see, e.g., Mallat (1998). This wavelet has the same time and frequency  $\rho$ pseudo support
$[-L,L]=[-3.5,3.5]$ with $\rho=0.9995$.
In the spectral domain, we have
\begin{equation}
\widehat{\psi} (t) = \hat{g}(\xi-\eta), \quad \hat{g}(\xi) = (4 \pi
\sigma^2)^{1/4} e^{\frac{-\sigma^2 \xi^2}{2}}
\end{equation}
We fit the Gabor wavelet inside the orthosympathetic or the parasympathetic
frequency bands by using Prop. \ref{Prop:scaling:modulation} or direct calculations. One
obtains the two Gabor wavelets  defined by (\ref{def:Gabor}) with the following
choice for the parameters:
\ban
\eta_1 = \frac{\omega_1+\omega_2}{2} & \mathrm{and}&  \sigma_1=\frac{2L}{\omega_2-\omega_1}\\
\eta_2 = \frac{\omega_2+\omega_3}{2} & \mathrm{and}&  \sigma_2=\frac{2L}{\omega_3-\omega_2}\\
\mathrm{moreover}\qquad\left|\rho\,pseudo\, Supp\, \psi_1\right|= \frac{4L^2}{\omega_2-\omega_1}& \mathrm{and}&\left|\rho\,pseudo\, Supp\, \psi_2\right|= \frac{4L^2}{\omega_3-\omega_2}\quad
\mathrm{with}\; \rho=0.9995\ean
One can see on Fig. 2 that the Fourier transforms $\widehat{\psi}_1(x)$ and
$\widehat{\psi}_2(x)$ still have  almost disjoint supports.
\end{itemize}

\begin{figure}[ht!]
\begin{center}
\includegraphics[height=7cm,width=16cm]{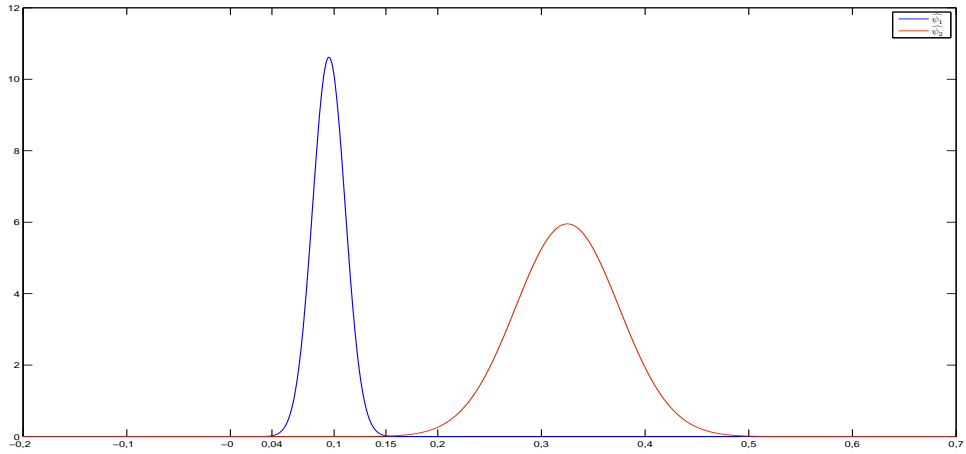}
\end{center}
\caption{The Fourier transforms
$\widehat{\psi}_1(x)$ (left) and $\widehat{\psi}_2(x)$ (right)}
\end{figure}

Using the Gabor wavelet is more efficient in terms of computing time, as it is
at least 8 times faster.  Figure 3 below displays the Gabor wavelets
coefficients in the orthosympathetic and parasympathetic bands respectively
for the sample plotted in Figure 1.

\begin{figure}[ht!]
\begin{center}
\includegraphics[height=8.5cm,width=8cm]{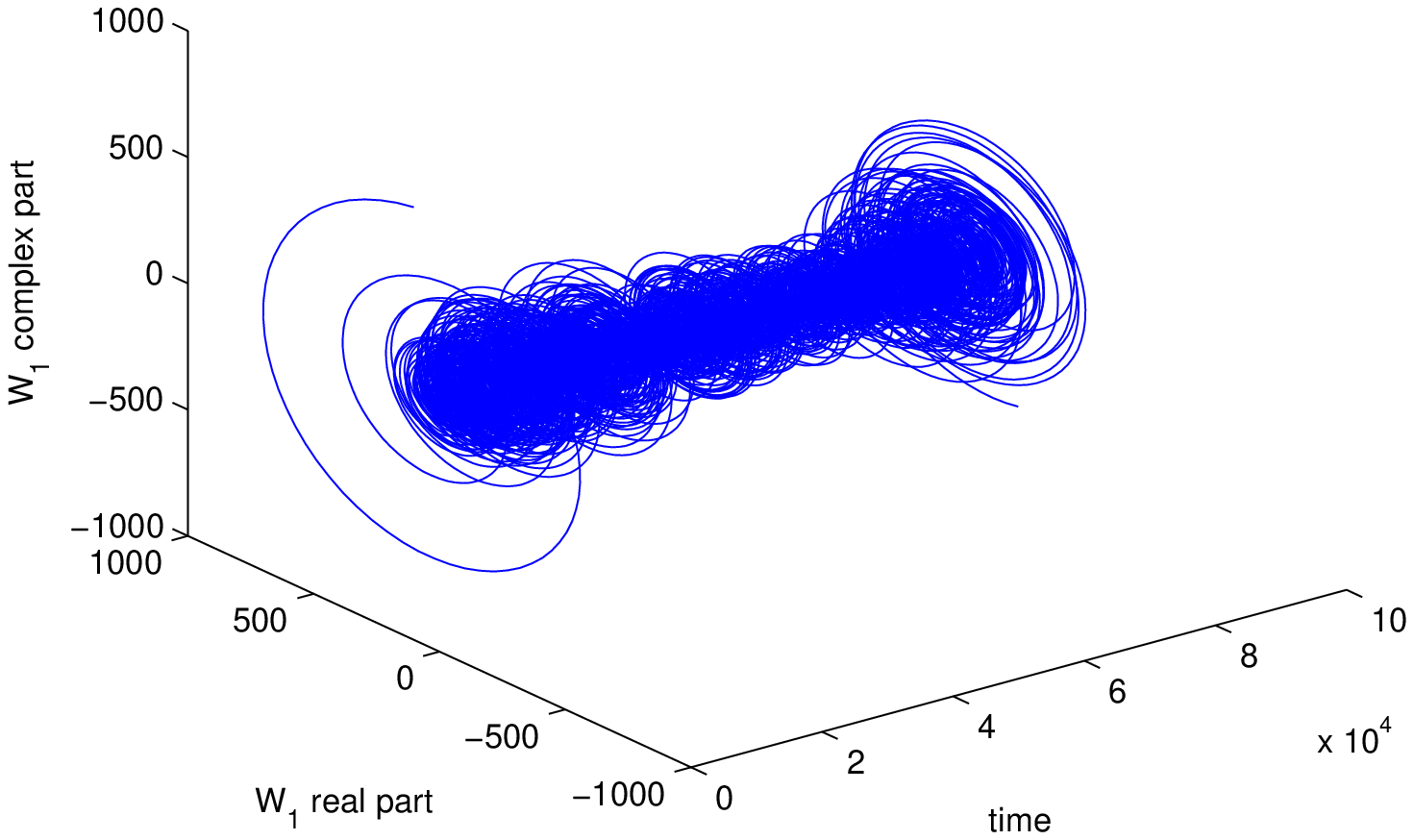}
\includegraphics[height=8.5cm,width=8cm]{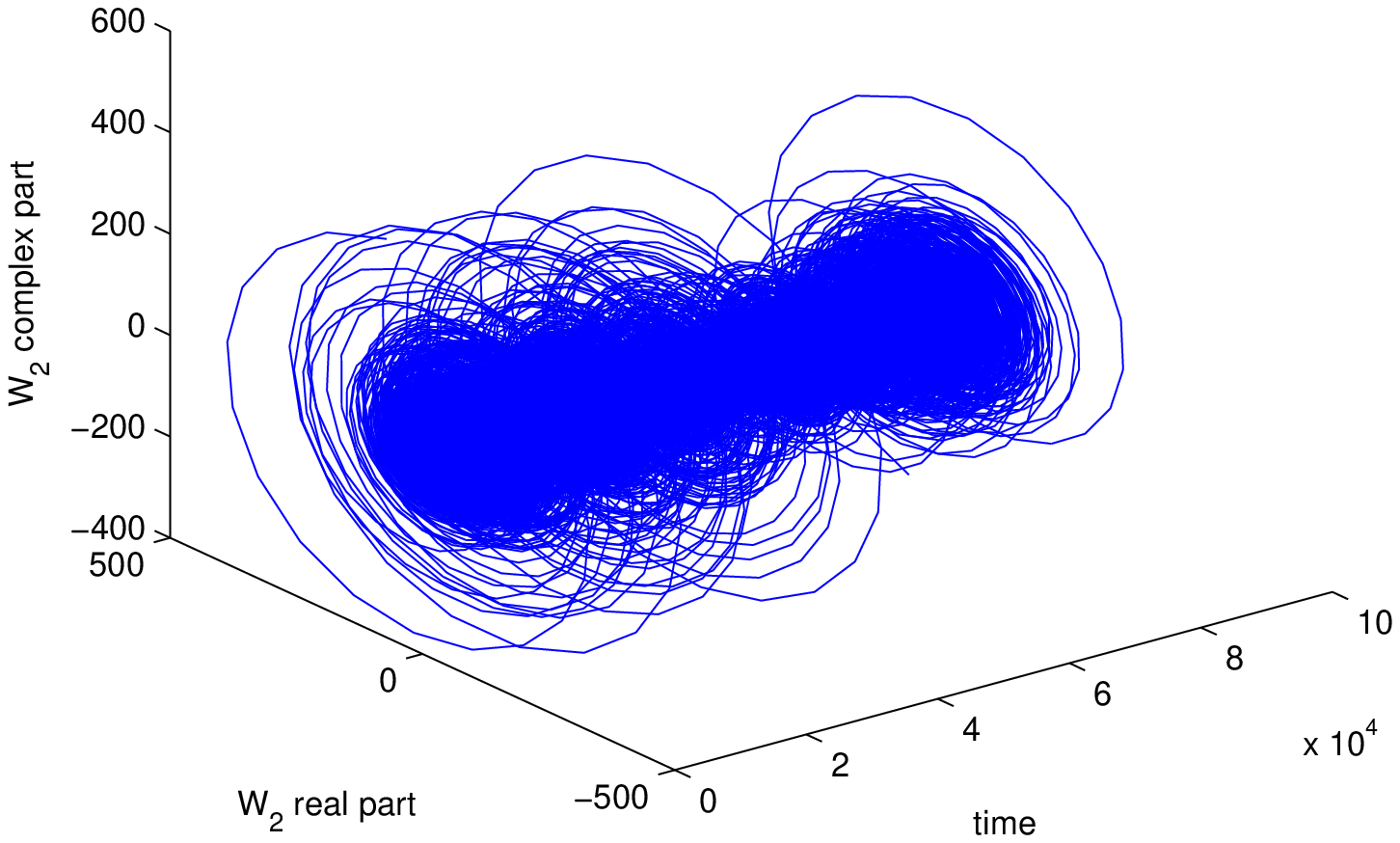}
\end{center}
\caption{The wavelet coefficients in the orthosympathetic band (left)
and in the parasympathetic band (right)
of the same healthy subject during a period of 24 hours}
\end{figure}

\section{Segmentation analysis}
We assume that the process $\{X_t\}$ is abruptly changing and is
characterized by a parameter $\theta \in \Theta$ that remains constant
between two changes. We  use this assumption to define
our contrast function $J(\bt,\bx)$.
Let $K$ be some integer and let $\bt =
\{\xt_1,\xt_2,\ldots,\xt_{K-1}\}$ be an ordered sequence of integers
satisfying $0<\xt_1<\xt_2<\ldots<\xt_{K-1}<n$.

For the detection of changes in the  mean and variance of a sequence of random
variables, i.e., a change in distribution, we use the following contrast
function, based on a Gaussian log--likelihood function:
\begin{equation}
\label{e:chg_dist}
J_n(\bt,\bx)=\frac{1}{n} \sum_{k=1}^{K} \frac{\| X_{\xt_{k}}
  -\bar{X}_{\xt_{k}}\|^2}{\hat{\sigma}_k^2}  + n_k \log  (\hat{\sigma}_k^2).
\end{equation}
where $n_k=\xt_k-\xt_{k-1}$ is the length of segment $k$,
$\hat{\sigma}_k^2$ is the empirical variance computed on that segment
$k$,
$\hat{\sigma}_k^2= n_k^{-1} \sum_{i=\xt_{k-1}+1}^{\xt_k}(X_i-\bar{X})^2$,
and $\bar{X}$ is the empirical mean of $X_1,\ldots,X_n$.

In fact, we minimize a penalized version of this contrast function, the
penalty parameter being selected in an adaptive way so that the obtained
segmentation does not depend too much on the penalty parameter; see also
Birg\'e and Massart (2007) for another choice for the penalty parameter.
Although this
method has been devised for Gaussian processes, it empirically provides
relevant results for non-Gaussian processes, e.g., financial time series,  see
Lavielle and Teyssi\`ere (2006) for further details.

\subsection{Orthosympathetic band}
For that frequency band, we obtain the following segmentation:
$$
\bt = \{28220, 33366, 71048\},$$ i.e.,
13h40'50'', 15h06'36'', 1h34'38''.


\begin{figure}[ht!]
\begin{center}
\includegraphics[height=6cm,width=16cm]{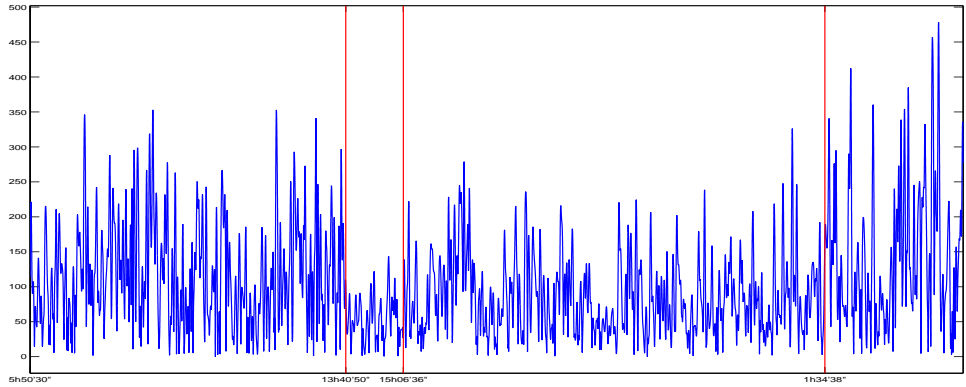}
\end{center}
\caption{Segmentation in the mean and variance (change in the distribution) of the modulus of the wavelet coefficients
  in the orthosympathetic band}
\end{figure}


\subsection{Parasympathetic band}
For that frequency band, we obtain the following segmentation:
$$
 \bt = \{11620, 21912, 28054, 31540, 36022,
  40172, 52622, 70550 \}, $$
i.e., 9h04'10'', 11h45'42'', 13h38'04'', 14h36'10'', 15h50'22'',
  17h00'02'', 20h27'32'', 1h26'40''.

\begin{figure}[ht!]
\begin{center}
\includegraphics[height=6cm,width=16cm]{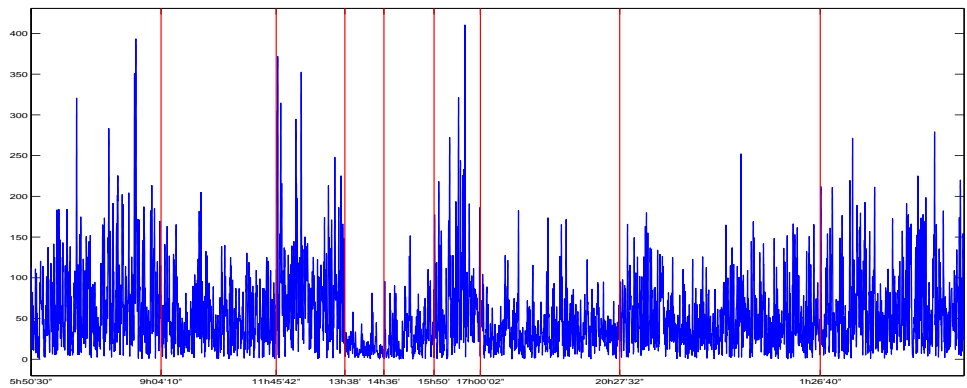}
\end{center}
\caption{Segmentation in the mean and variance (change in the distribution) of the modulus of the wavelet
  coefficients in the parasympathetic band}
\end{figure}

\section{Conclusion}
The example illustrated by Figures 4, 5 and 6 shows at around $t=28,220$ a simultaneous variation of
both systems. But, one can also
observe  change-points in the  orthosympathetic and parasympathetic
bands occurring at different times. In the future, we will study the existence of
a possible causality or sequentiality between these change--points in
different bands.

\begin{figure}[ht!]
\begin{center}
\includegraphics[height=6cm,width=16cm]{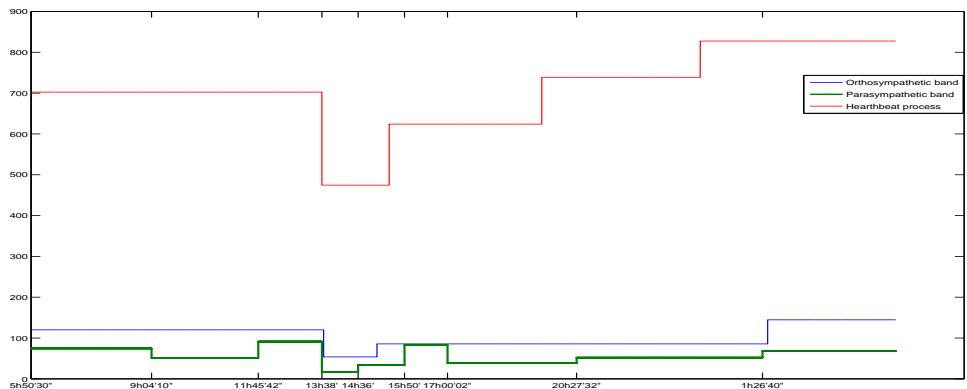}
\end{center}
\caption{Mean of the modulus if the wavelets coefficients for the
  orthosympathetic and parasympathetic bands, and of the RR process
  $X(t)$}
\end{figure}

\end{document}